\documentclass[11pt]{article}
\usepackage{setspace}
\usepackage[utf8]{inputenc}
\usepackage{amsfonts,url,epsfig}
\usepackage[colorlinks,linkcolor=black,citecolor=black,urlcolor=black,filecolor=black,backref=false]{hyperref}
\usepackage{geometry} 
\usepackage{sectsty} 
\usepackage[normalem]{ulem}
\usepackage{microtype}

\usepackage{todonotes}

\sectionfont{\sffamily\bfseries\upshape\large} 
\subsectionfont{\sffamily\bfseries\upshape\normalsize} 
\subsubsectionfont{\sffamily\mdseries\upshape\normalsize}
\makeatletter
\renewcommand\@seccntformat[1]{\csname the#1\endcsname.\quad}
\makeatother\renewcommand{\bibitem}{\vskip2pt\par\hangindent\parindent\hskip-\parindent}

\makeatletter
\def\@maketitle{%
  \begin{center}%
  \let \footnote \thanks
    {\large \@title \par}%
    {\normalsize
      \begin{tabular}[t]{c}%
        \@author
      \end{tabular}\par}%
    {\small \@date}%
  \end{center}%
}
\makeatother

\title{\bf What are the most important statistical ideas of the past 50 years?\footnote{To appear in the {\em Journal of the American Statistical Association}.  We thank Bin Yu, Brad Efron, Tom Belin, Trivellore Raghunathan, Chuanhai Liu, Sander Greenland, Howard Wainer, Anna Menacher, and anonymous reviewers for helpful comments.}
\vspace{.1in}}

\author{Andrew Gelman\footnote{Department of Statistics and Department of Political Science, Columbia University, New York.} \ and Aki Vehtari\footnote{Department of Computer Science, Aalto University, Espoo, Finland.} \vspace{.1in}}

\date{3 June 2021}

\begin{document}

\maketitle


\begin{abstract}
We review the most important statistical ideas of the past half century, which we categorize as:  counterfactual causal inference, bootstrapping and simulation-based inference, overparameterized models and regularization, Bayesian multilevel models, generic computation algorithms, adaptive decision analysis, robust inference, and exploratory data analysis.  We discuss key contributions in these subfields, how they relate to modern computing and big data, and how they might be developed and extended in future decades.  The goal of this article is to provoke thought and discussion regarding the larger themes of research in statistics and data science.
\end{abstract}

\section{The most important statistical ideas of the past 50 years}\label{important}

A lot has happened in the past half century! The eight ideas reviewed below represent a categorization based on our experiences and reading of the literature and are not listed in chronological order or in order of importance.  They are separate concepts capturing different useful and general developments in statistics.  The present review is intended to cover the territory and is influenced not just by our own experiences but also by discussions with others; nonetheless we recognize that any short overview will be incomplete, and we welcome further discussions from other perspectives.

Each of these ideas has pre-1970 antecedents, both in the theoretical statistics literature and in the practice of various applied fields.  But each has developed enough in the past fifty years to have become something new.

\subsection{Counterfactual causal inference}

We begin with a cluster of different ideas that have appeared in statistics, econometrics, psychometrics, epidemiology, and computer science, all revolving around the challenges of causal inference, and all in some way bridging the gap between, on one hand, naive causal interpretation of observational inferences and, on the other, the recognition that correlation does not imply causation.  The key idea is that causal identification is possible, under assumptions, and that one can state these assumptions rigorously and address them, in various ways, through design and analysis.  Debate continues on the specifics of how to apply causal models to real data, but the work in this area over the past fifty years has allowed much more precision on the assumptions required for causal inference, and this in turn has stimulated work in statistical methods for these problems.

Different methods for causal inference have developed in different fields.  In econometrics the focus has been on the structural models and their implications for average treatment effects (Imbens and Angrist, 1994), in epidemiology the focus has been on inference with observational data (Greenland and Robins, 1986), psychologists have been aware of the importance of interactions and varying treatment effects (Cronbach, 1975), in statistics there has been work on matching and other approaches to adjust for and measure differences between treatment and control groups (Rosenbaum and Rubin, 1983).  In all this work there has been a common thread of modeling causal questions in terms of counterfactuals or potential outcomes, which is a big step beyond the earlier standard approach which did not clearly distinguish between descriptive and causal inferences.  Key developments include Neyman (1923), Welch (1937), Rubin (1974), and Haavelmo (1943); see Heckman and Pinto (2015) for some background and VanderWeele (2015) for a recent review. 

The purpose of the aforementioned methods is to define and estimate the effect of some specified treatment or exposure, adjusting for biases arising from imbalance, selection, and measurement errors.  Another important area of research has been in causal discovery, where the goal is not to estimate a particular treatment effect but rather to learn something about the causal relations among several variables.  There is a long history of such ideas using methods of path analysis, from researchers in various fields of application such as genetics (Wright, 1923), economics (Wold, 1954), and sociology (Duncan, 1975); as discussed by Wermouth (1980), these can be framed in terms of simultaneous equation models. Influential recent work in this area has linked to probabilistic ideas of graphical models (Spirtes, Glymour, and Scheines, 1993, Heckerman, Geiger, and Chickering, 1995, Peters, Janzing, and Schölkopf, 2017). An important connection to psychology and computer science has arisen based on the idea that causal identification is a central task of cognition and thus should be a computable problem that can be formalized mathematically (Pearl, 2009).  Path analysis and causal discovery can be framed in terms of potential outcomes, and vice versa (Morgan and Winship, 2014).  However formulated, ideas and methods of counterfactual reasoning and causal structure have been influential within statistics and computer science and also in applied research and policy analysis.

\subsection{Bootstrapping and simulation-based inference}

A trend of statistics in the past fifty years has been the substitution of computing for mathematical analysis, a move that began even before the onset of ``big data'' analysis.  Perhaps the purest example of a computationally defined statistical method is the bootstrap, in which some estimator is defined and applied to a set of randomly resampled datasets (Efron, 1979, Efron and Tibshirani, 1993).  The idea is to consider the estimate as an approximate sufficient statistic of the data and to consider the bootstrap distribution as an approximation to the sampling distribution of the data. At a conceptual level, there is an appeal to thinking of prediction and resampling as fundamental principles from which one can derive statistical operations such as bias correction and shrinkage (Geisser, 1975).

Antecedents include the jackknife and cross validation (Quenouille, 1949, Tukey, 1958, Stone, 1974, Geisser, 1975), but there was something particularly influential about the bootstrap idea in that its generality and simple computational implementation allowed it to be immediately applied to a wide variety of applications where conventional analytic approximations failed; see for example Felsenstein (1985). Availability of sufficient computational resources also helped as it became trivial to repeat inferences for many resampled datasets.

The increase in computational resources has made other related resampling and simulation based approaches popular as well.
In permutation testing, resampled datasets are generated by breaking the (possible) dependency between the predictors and target by randomly shuffling the target values. Parametric bootstrapping, prior and posterior predictive checking (Box, 1980, Rubin, 1984), and simulation-based calibration 
all create replicated datasets from a model instead of directly resampling from the data.  Sampling from a known data generating mechanism is commonly used to create simulation experiments to complement or replace mathematical theory when analyzing complex models or algorithms.

\subsection{Overparameterized models and regularization}

A major change in statistics since the 1970s, coming from many different directions, is the idea of fitting a model with a large number of parameters---sometimes more parameters than data points---using some regularization procedure to get stable estimates and good predictions.  The idea is to get the flexibility of a nonparametric or highly parameterized approach, while avoiding the overfitting problem.  Regularization can be implemented as a penalty function on the parameters or on the predicted curve (Good and Gaskins, 1971).

Early examples of richly parameterized models include Markov random fields (Besag, 1974), splines (Wahba and Wold, 1975, Wahba, 1978), and Gaussian processes (O’Hagan, 1978), followed by classification and regression trees (Breiman et al., 1984), neural networks (Werbos, 1981, Rumelhart, Hinton, and Williams, 1987, Buntine and Weigend, 1991, MacKay, 1992, Neal, 1996), wavelet shrinkage (Donoho and Johnstone, 1994), lasso, horseshoe, and other alternatives to least squares (Dempster, Schatzoff, and Wermuth, 1977, Tibshirani, 1996, Carvalho, Polson, and Scott, 2010), and support vector machines (Cortes and Vapnik, 1995) and related theory (Vapnik, 1998).

The 1970s also saw the start of the development of Bayesian nonparametric priors on infinite dimensional families of probability models (Müller and Mitra, 2013), such as Dirichlet processes (Ferguson, 1973), Chinese restaurant processes (Aldous, 1985), Polya trees (Lavine, 1992, Mauldin et al., 1992) and Pitman and Yor (1997) processes, and many other examples since then
All these models have the feature of expanding with sample size, and with parameters that did not always have a direct interpretation but rather were part of a larger predictive system. In the Bayesian approach the prior could be first considered in a function space, with the corresponding prior for the model parameters then derived indirectly.

Many of these models had limited usage until enough computational resources became easily available.
Overparameterized models have continued to be developed in image recognition (Wu et al., 2004) and deep neural nets (Bengio, LeCun, and Hinton, 2015, Schmidhuber, 2015). Hastie, Tibshirani, and Wainwright (2015) have framed much of this work as the estimation of sparse structure, but we view regularization as being more general in that it also allows for dense models to be fit to the extent supported by data.  

Along with a proliferation of statistical methods and their application to larger datasets, researchers have developed methods for tuning, adapting, and combining inferences from multiple fits, including stacking (Wolpert, 1992), Bayesian model averaging (Hoeting et al., 1999), boosting (Freund and Schapire, 1997), and gradient boosting (Friedman, 2001).  These advances have been accompanied by an alternative view of the foundations of statistics based on prediction rather than modeling (Breiman, 2001).

\subsection{Bayesian multilevel models}

Multilevel or hierarchical models have parameters that vary by group, allowing models to adapt to cluster sampling, longitudinal studies, time-series cross-sectional data, meta-analysis, and other structured settings. In a regression context, a multilevel model can be viewed as a particular parameterized covariance structure or as a probability distribution where the number of parameters increases in proportion to the data.

Multilevel models can be seen as Bayesian in that they include probability distributions for unknown latent characteristics or varying parameters.  Conversely, Bayesian models have a multilevel structure with distributions for data given parameters and for parameters given hyperparameters.

The idea of partial pooling of local and general information is inherent in the mathematics of prediction from noisy data and, as such, dates back to Laplace and Gauss and is implicit in the ideas of Galton.  Partial pooling was used in specific application areas such as animal breeding (Henderson et al., 1959), and its general relevance to multiplicity in statistical estimation problems was given a theoretical boost by the work of Stein (1955) and James and Stein (1960), ultimately inspiring work in areas ranging from psychology (Novick et al., 1972) to pharmacology (Sheiner, Rosenberg, and Melmon, 1972) to survey sampling (Fay and Herriot, 1979).  Lindley and Smith (1972) and Lindley and Novick (1981) supplied a mathematical structure based on estimating hyperparameters of the multivariate normal distribution, with Efron and Morris (1971, 1972) providing a corresponding decision-theoretic justification, and then these ideas were folded into regression modeling and applied to a wide range of problems with structured data (for example, Liang and Zeger, 1986, and Lax and Phillips, 2012).  From a different direction, shrinkage of multivariate parameters has been given an information-theoretic justification (Donoho, 1995).
Rather than considering multilevel modeling as a specific statistical model or computational procedure, we prefer to think of it as a framework for combining different sources of information, and as such it arises whenever we wish to make inferences from a subset of data (small-area estimation) or to generalize data to new problems (meta-analysis).  Similarly, Bayesian inference has been valuable not just as a way of combining prior information with data but also as a way of accounting for uncertainty for inference and decision making.

\subsection{Generic computation algorithms}

The advances in modeling we have discussed have only become possible due to modern computing.  But this is not just larger memory, faster CPUs, efficient matrix computations, user-friendly languages, and other innovations in computing.  A key component has been advances in statistical algorithms for efficient computing.

The innovative statistical algorithms of the past fifty years are statistical in the sense of being motivated and developed in the context of the structure of a statistical problem.  The EM algorithm (Dempster, Laird, and Rubin, 1977, Meng and van Dyk, 1997), Gibbs sampler (Geman and Geman, 1984, Gelfand and Smith, 1990), particle filters (Kitagawa, 1993, Gordon et al., 1993, Del Moral, 1996), variational inference (Jordan et al., 1999), and expectation propagation (Minka, 2001, Heskes et al., 2005) in different ways make use of the conditional independence structures of statistical models.  The Metropolis algorithm (Hastings, 1970) and hybrid or Hamiltonian Monte Carlo (Duane et al., 1987) were less directly motivated by statistical concerns---these were methods that were originally developed to compute high-dimensional probability distributions in physics---but they have become adapted to statistical computing in the same way that optimization algorithms were adopted in an earlier era to compute least squares and maximum likelihood estimates.
The method called approximate Bayesian computation, in which posterior inferences are obtained by simulating from the generative model instead of evaluating the likelihood function, can be useful if the analytic form of the likelihood is intractable or very costly to compute (Rubin, 1984, Tavaré et al., 1997, Marin et al., 2012). Martin, Frazier, and Robert (2020) review the history of computational methods in Bayesian statistics.

Throughout the history of statistics, advances in data analysis, probability modeling, and computing have gone together, with new models motivating innovative computational algorithms and new computing techniques opening the door to more complex models and new inferential ideas, as we have already noted in the context of high-dimensional regularization, multilevel modeling, and the bootstrap. The generic automatic inference algorithms allowed decoupling the development of the models so that changing the model did not require changes to the algorithm implementation.

\subsection{Adaptive decision analysis}
From the 1940s through the 1960s, decision theory was often framed as foundational to statistics, via utility maximization (Wald, 1949, Savage, 1954), error-rate control (Tukey, 1953, Scheffé, 1959), and empirical Bayes analysis (Robbins, 1959, 1964), and recent decades have seen developments following up this work, in Bayesian decision theory (Berger, 1985) and false discovery rate analysis (Benjamini and Hochberg, 1995).  Decision theory has also been influenced from the outside by psychology research on heuristics and biases in human decision making (Kahneman, Slovic, and Tversky, 1982, Gigerenzer and Todd, 1999).

One can also view decision making as an area of statistical application. Some important developments in statistical decision analysis involve Bayesian optimization (Mockus, 1974, 2012, Shariari et al., 2015) and reinforcement learning (Sutton and Barto, 2018), which are related to a renaissance in experimental design for A/B testing in industry and online learning in many engineering applications. Recent advances in computation have made it possible to use richly parameterized models such as Gaussian process and neural networks as priors for functions in adaptive decision analysis, and to perform large-scale reinforcement learning, for example to create artificial intelligence to control robots, generate text, and play games such as Go (Silver et al., 2017).

Much of this work has been done outside of statistics, with methods such as nonnegative matrix factorization (Paatero and Tapper, 1994), nonlinear dimension reduction (Lee and Verleysen, 2007), generative adversarial networks (Goodfellow et al., 2014), and autoencoders (Goodfellow, Bengio, and Courville, 2016):  these are all unsupervised learning
methods for finding structures and decompositions.

\subsection{Robust inference}

The idea of robustness is central to modern statistics, and it's all about the idea that we can use models even when they have assumptions that are not true.  An important part of statistical theory is to develop models that work well, under realistic violations of these assumptions.  Early work in this area was synthesized by Tukey (1960); see Stigler (2010) for a historical review.  Following the theoretical work of Huber (1972) and others, researchers have developed robust methods that have been influential in practice, especially in economics, where there is acute awareness of the imperfections of statistical models.  In economic theory there is the idea of the ``as if'' analysis and the reduced-form model, so it makes sense that econometricians are interested in statistical procedures that work well under a range of assumptions.  For example, applied researchers in economics and other social sciences make extensive use of robust standard errors (White, 1980) and partial identification (Manski, 1990).

In general, though, the main impact of robustness in statistical research is not in the development of particular methods, so much as in the idea of evaluating statistical procedures under what Bernardo and Smith (1994) call the $\mathcal{M}$-open world in which the data-generating process does not fall within the class of fitted probability models.  Greenland (2005) has argued that researchers should explicitly account for sources of error that are not traditionally included in statistical models. Concerns of robustness are relevant for the densely parameterized models that are characteristic of much of modern statistics, and this has implications for model evaluation more generally (Navarro, 2018).  There is a connection between robustness of a statistical method to model misspecification, and a workflow involving model checking and model improvement (Box, 1980).

\subsection{Exploratory data analysis}

The statistical ideas discussed above all involve some mixture of intense theory and intense computation.  From a completely different direction, there has been an influential back-to-basics movement, eschewing probability models and focusing on graphical visualization of data.  The virtues of statistical graphics were convincingly argued in influential books by Tukey (1977) and Tufte (1983), and many of these ideas entered statistical practice through their implementation in the data analysis environment {\textsf S} (Chambers et al., 1983), a precursor to {\textsf R}, which is currently the dominant statistics software in many areas of statistics and its application.

Following Tukey (1962), the proponents of exploratory data analysis have emphasized the limitations of asymptotic theory and the corresponding benefits of open-ended exploration and communication (Cleveland, 1985) along with a general view of data science as going beyond statistical theory (Chambers, 1993, Donoho, 2017).  This fits into a view of statistical modeling that is focused more on discovery than on the testing of fixed hypotheses, and as such has been influential not just in the development of specific graphical methods but also in moving the field of statistics away from theorem-proving and toward a more open and, we would say, healthier perspective on the role of learning from data in science.  An example in medical statistics is the much-cited paper by Bland and Altman (1986) that recommends graphical methods for data comparison in place of correlations and regressions.

In addition, attempts have been made to formalize exploratory data analysis:  Gelman (2003) connects data display and visualization to Bayesian predictive checks, and Wilkinson (2005) formalizes the comparisons and data structures inherent in statistical graphics, in a way that Wickham (2016) was able to implement into a highly influential set of {\textsf R} packages that has transformed statistical practice in many fields.

Advances in computation have allowed practitioners to build large complicated models quickly, leading to a process in which ideas of statistical graphics are useful in understanding the relation between data, fitted model, and predictions.  The term ``exploratory model analysis'' (Unwin, Volinsky, and Winkler, 2003, Wickham, 2006) has sometimes been used to capture the experimental nature of the data analysis process, and efforts have been made to include visualization within the workflow of model building and data analysis (Gabry et al., 2019, Gelman et al., 2020). 

\section{What these ideas have in common and how they differ}

It would be tempting to say that a common feature of all these methods is catchy names and good marketing.  But we suspect that the names of these methods are catchy only in retrospect.  Terms such as ``counterfactual,'' ``bootstrap,''  ``stacking,'' and ``boosting'' could well sound jargony rather than impressive, and we suspect it is the value of the methods that has made the names sound appealing, rather than the reverse. 

\subsection{Ideas lead to methods and workflows}\label{opening}

The benefit of application to statistical theory is clear.  What about the benefits the other way?  Most directly, one can view theory as a shortcut to computation.  Such shortcuts will always be needed:  demands for modeling inevitably grow with computing power, hence the value of analytic summaries and approximations.  In addition, theory can help us understand how a statistical method works, and the logic of mathematics can inspire new models and approaches to data analysis.

We consider the ideas listed above to be particularly important in that each of them was not so much a method for solving an existing problem, as an opening to new ways of thinking about statistics and new ways of data analysis.

To put it another way, each of these ideas was a codification, bringing inside the tent an approach that had been considered more a matter of taste or philosophy than statistics:
\begin{itemize}
\item The counterfactual framework placed causal inference within a statistical or predictive framework in which causal estimands could be precisely defined and expressed in terms of unobserved data within a statistical model, connecting to ideas in survey sampling and missing-data imputation (Little, 1993, Little and Rubin, 2002).
\item The bootstrap opened the door to a form of implicit nonparametric modeling.
\item Overparameterized models and regularization formalized and generalized the existing practice of restricting a model's size based on the ability to estimate its parameters from the data, which is related to cross validation and information criteria (Akaike, 1973, Mallows, 1973, Watanabe, 2010).
\item Multilevel models formalized ``empirical Bayes'' techniques of estimating a prior distribution from data, leading to the use of such methods with more computational and inferential stability in a much wider class of problems.
\item Generic computation algorithms make it possible for applied practitioners to quickly fit advanced models for causal inference, multilevel analysis, reinforcement learning, and many other areas, leading to a broader impact of core ideas in statistics and machine learning.
\item Adaptive decision analysis connects engineering problems of optimal control to the field of statistical learning, going far beyond classical experimental design.
\item Robust inference formalized intuitions about inferential stability, framing these questions in a way that allowed formal evaluation and modeling of different procedures to handle otherwise nebulous concerns about outliers and model misspecification, and ideas of robust inference have informed ideas of nonparametric estimation (Owen, 1988).
\item Exploratory data analysis moved graphical techniques and discovery into the mainstream of statistical practice, just in time for the use of these tools to better understand and diagnose problems of new complex classes of probability models that are being fit to data.
\end{itemize}

\subsection{Advances in computing}

Meta-algorithms---workflows that make use of existing models and inferential procedures---have always been with us in statistics:  consider least squares, the method of moments, maximum likelihood, and so forth. One characteristic aspect of many of the machine learning meta-algorithms that have been developed in the past fifty years is that they involve splitting the data or model in some way.  The learning meta-algorithms are associated with divide-and-conquer computational methods, most notably variational Bayes and expectation propagation, which can be viewed as generalizations of algorithms that iterate over parameters or that combine inference from subsets of the data.

Meta-algorithms and iterative computations are an important development in statistics for two reasons. First, the general idea of combining information from multiple sources, or creating a strong learner by combining weak learners, can be applied broadly, beyond the examples where such meta-algorithms were originally developed.  Second, adaptive algorithms play well with online learning and ultimately can be viewed as representing a modern view of statistics in which data and computation are dispersed, a view in which information exchange and computational architecture are part of the meta-model or inferential procedure (Efron and Hastie, 2016).

It is no surprise that new methods take advantage of new technical tools:  as computing improves in speed and scope, statisticians are no longer limited to simple models with analytic solutions and simple closed-form algorithms such as least squares.  We can outline how the above-listed ideas make use of modern computation:
\begin{itemize}
\item Several of the ideas---bootstrapping, overparameterized models, and machine learning meta-analysis---directly take advantage of computing speed and could not easily be imagined in a pre-computer world. For example, the popularity of neural networks increased substantially only after the introduction of efficient GPU cards and cloud computing.
\item Also important, beyond computing power, is the dispersion of computing resources: desktop computers allowed statisticians and computer scientists to experiment with new methods and then allowed practitioners to use them.
\item Exploratory data analysis began with pencil-and-paper graphs but has completely changed with developments in computer graphics.
\item In the past, Bayesian inference was constrained to simple models that could be solved analytically. With the increase in computing power, variational and Markov chain simulation methods have allowed separation of model building and development of inference algorithms, leading to probabilistic programming that has freed domain experts in different fields to focus on model building and get inference done automatically. This resulted in an increase in popularity of Bayesian methods in many applied fields starting in the 1990s.
\item Adaptive decision analysis, Bayesian optimization, and online learning are used in computationally and data-intensive problems such as optimizing big machine learning and neural network models, real-time image processing, and natural language processing.
\item Robust statistics are not necessarily computationally intensive, but their use was associated with a computation-fueled move away from closed-form estimates such as least squares.  The development and understanding of robust methods was facilitated by a simulation study that used extensive computation for its time (Andrews et al., 1972).
\item Shrinkage for multivariate inference can be justified not just by statistical efficiency but also on computational grounds, motivating a new kind of asymptotic theory (Donoho, 2006, Candès, Romberg, and Tao, 2008).
\item The key ideas of counterfactual causal inference are theoretical, not computational, but in recent years causal inference has advanced by the use of computationally intensive nonparametric methods, leading to a unification of causal and predictive modeling in statistics, economics, and machine learning (Hill, 2011, Wager and Athey, 2018, Chernozhukov et al., 2018).
\end{itemize}

\subsection{Big data}
In addition to the opportunities opened up for statistical analysis, modern computing has also yielded big data in ways that have inspired the  application and development of new statistical methods:  examples include gene arrays, streaming image and text data, and online control problems such as self-driving cars.  Indeed, one reason for the popularity of the term ``data science'' is because, in such problems, data processing and efficient computing can be as important as the statistical methods used to fit the data.

A common feature of all the ideas discussed in this paper and they facilitate the use of more data, compared to previously existing approaches:
\begin{itemize}
\item The counterfactual framework allows causal inference from observational data using the same structure used to model controlled experiments.
\item Bootstrapping can be used for bias correction and variance estimation for complex surveys, experimental designs, and other data structures where analytical calculations are not possible.
\item Regularization allows users to include more predictors in a model without so much concern about overfitting.
\item Multilevel models use partial pooling to incorporate information from different sources, applying the principle of meta-analysis more generally.
\item Generic computation algorithms allow users to fit larger models, which can be necessary to connect available data to underlying questions of interest.
\item Adaptive decision analysis makes use of stochastic optimization methods developed in numerical analysis.
\item Robust inference allows more routine use of data with outliers, correlations, and other aspects that could get in the way of conventional statistical modeling.
\item Exploratory data analysis opens the door to visualization of complex datasets and has motivated the development of tidy data analysis and the integration of statistical analysis, computation, and communication.
\end{itemize}

The past fifty years have also seen the development of statistical programming environments, most notably {\textsf S} (Becker, Chambers, and Wilks, 1988) and then {\textsf R} (Ihaka and Gentleman, 1996), and general-purpose inference engines beginning with BUGS (Spiegelhalter et al., 1994) and its successors (Lunn et al., 2009).  More recently, ideas of numerical analysis, automated inference, and statistical computing have started to mix, in the form of reproducible research environments such as Jupyter notebooks and probabilistic programming environments such as Stan, Tensorflow, and Pyro (Stan Development Team, 2020, Tensorflow, 2020, Pyro, 2020).  So we can expect at least some partial unification of inferential and computing methods, as demonstrated for example by the use of automatic differentiation for optimization, sampling, and sensitivity analysis.

\subsection{Connections and interactions among these ideas}

Stigler (2016) has argued for the relevance of certain common themes underlying apparently disparate areas of statistics.  This idea of interconnection can be seen to apply to recent developments as well.
For example, what is the connection between robust statistics (which focuses on departures from particular model assumptions) and exploratory data analysis (which is traditionally presented as being not interested in models at all)?  Exploratory methods such as residual plots and hanging rootograms can be derived from specific model classes (additive regression and the Poisson distribution, respectively) but their value comes in large part from their interpretability without reference to the models that inspired them.  One can similarly consider a method such as least squares on its own terms, as an operation on data, then study the class of data generating processes for which it will perform well, and then use the results of such a theoretical analysis to propose more robust procedures that extend the range of useful applicability, whether defined based on breakdown point, minimax risk, or otherwise.  Conversely, purely computational methods such as Monte Carlo evaluation of integrals can fruitfully be interpreted as solutions to statistical inference problems (Kong et al., 2003).

For another connection, the potential outcome framework for causal inference, which allows a different treatment effect for each unit in the population, lends itself naturally to a meta-analytic approach in which effects can vary, and this can be modeled using multilevel regression in the analyses of experiments or observational studies.  Work on the bootstrap can, in retrospect, give us a new perspective on empirical Bayes (multilevel) inference as a nonparametric approach in which a normal distribution or other parametric model is used for partial pooling but final estimates are not restricted to any parametric form.  And research on regularizing wavelets and other richly parameterized models has an unexpected connection to the stable inferential procedures developed in the context of robustness.

Other methodological connections are more obvious.  Regularized overparameterized models are optimized using machine-learning meta-algorithms, which in turn can yield inferences that are robust to contamination. To draw these connections another way, robust regression models correspond to mixture distributions which can be viewed as multilevel models, and these can be fitted using Bayesian inference.  Deep learning models are related to a form of multilevel logistic regression and relates to reproducing kernel Hilbert spaces, which are used in splines and support vector machines (Kimeldorf and Wahba, 1971, Wahba, 2002).

Highly parameterized machine learning methods can be framed as Bayesian hierarchical models, with regularizing penalty functions corresponding to hyperpriors, and unsupervised learning models can be framed as mixture models with unknown group memberships.  In many cases the choice of whether to use a Bayesian generative framework depends on computation, and this can go in both ways:  Bayesian computational methods can help capture uncertainty in inference and prediction, and efficient optimization algorithms can be used to approximate model-based inference.

Many of the ideas we have been discussing involve rich parameterizations followed by some statistical or computational tools for regularization.  As such, they can be considered as more general implementations of the idea of sieves---models that get larger as more data become available (Grenander, 1981, Geman and Hwang, 1982, Shen and Wong, 1994).

\subsection{Links to other new and useful developments in statistics}

Where do particular statistical models fit into our story?  Here we are thinking of influential work such as hazard regression (Cox, 1972), generalized linear models (Nelder, 1977, McCullagh and Nelder, 1989), structural equation models (Baron and Kenny, 1986), latent classification (Blei, Ng, and Jordan, 2003), Gaussian processes (O'Hagan, 1978, Rasmussen and Williams, 2006), and deep learning (Hinton, Osindero, and Teh, 2006, Bengio, LeCun, and Hinton, 2015, Schmidhuber, 2015), and models for structured data such as time series (Box and Jenkins, 1976, Brillinger, 1981), spatial processes (Besag, 1974, 1986), network data (Kolaczyk, 2009), and self-similar processes (Künsch, 1987).  These models and their associated applied successes can be thought of demonstrations of the ideas developed in the first section of this article, or as challenges that motivated many of these developments (for example, generalized linear models with many predictors motivating regularization methods, or Gaussian process models motivating advances in approximate computation and a shift toward predictive evaluation), or as bridges between different statistical ideas (for example, structural equation models connecting graphical models and causal inference, or deep learning connecting Bayesian multilevel models and generic computation algorithms).  It is not possible to disentangle models, methods, applications, and principles, and alternative histories of statistics take any of these as an organizing principle.

To discuss the connections among different conceptual advances is not to deny that debates remain regarding appropriate use and interpretation of statistical methods.  For example, there is a duality between false discovery rate and multilevel modeling, but procedures based on these different principles can give different results.  Multilevel models are typically fit using Bayesian methods, and nothing is pooled all the way to zero in the posterior distribution.  In contrast, false discovery rate methods are typically applied using $p$-value thresholds, with the goal of identifying some small number of statistically significantly nonzero results.  For another example, in causal inference, there is increasing interest in densely-parameterized machine learning predictions followed by poststratification to obtain population causal estimates of specified exposures or treatments, but in more open-ended settings there is the goal of discovering nonzero causal relationships.  Again, different methods are used, depending on whether the aim is dense prediction or sparse discovery.  


Finally, we can connect research in statistical methods to trends in the application of statistics within science and engineering.  An entire series of articles could be written just on this topic.  Here we mention one such area, the replication crisis or reproducibility revolution in biology, psychology, economics, and other sciences. Landmark papers in the reproducibility revolution include Meehl (1978) outlining the philosophical flaws in the standard use of null hypothesis significance testing to make scientific claims, Ioannidis (2005) arguing that most published studies in medicine were making claims unsupported by their statistical data, and Simmons, Nelson, and Simonsohn (2011) explaining how ``researcher degrees of freedom'' can enable researchers to routinely obtain statistical significance even from data that are pure noise.  Some of the proposed remedies are procedural (for example, Amrhein, Greenland, and McShane, 2019), but there have also been suggestions that some of the problems with nonreplicable research can be resolved using multilevel models, partially pooling estimates toward zero to better reflect the population of effect sizes under study (van Zwet, Schwab, and Senn, 2020).
Questions of reproducibility and stability also relate directly to bootstrapping and robust statistics (Yu, 2013).

\section{What will be the important statistical ideas of the next few decades?}

\subsection{Looking backward}

In considering the most important developments since 1970, it could also make sense to reflect upon
the most important statistical ideas of 1920--1970 (these could include quality control, latent-variable modeling, sampling theory, experimental design, classical and Bayesian decision analysis, confidence intervals and hypothesis testing, maximum likelihood, the analysis of variance, and objective Bayesian inference---quite a list!), 1870--1920 (classification of probability distributions, regression to the mean, phenomenological modeling of data), and previous centuries, as studied by Stigler (1986) and others.

In this article we have attempted to offer a broad perspective, reflecting the different perspectives of the authors.  But others will have their own takes on what are the most important statistical ideas of the past fifty years, and another view is gained by looking at the topics of papers published in statistics journals (Anderlucci, Montanari, and Viroli, 2019).  Indeed, the point of asking what are the most important ideas is not so much to answer the question, as to stimulate discussion of what it means for a statistical idea to be important. In the present article, we have avoided ranking papers by citation counts or other numerical measure, but implicitly we are measuring intellectual influence in a page-rank-like way, in that we are trying to focus on the ideas that have influenced the development of methods that have influenced statistical practice.

We are interested in others' views on what are the most influential statistical ideas of the last half century and how these ideas have combined to affect the practice of statistics and scientific learning.

\subsection{Looking forward}

What will come next?  We agree with Popper (1957) that one can't anticipate all future scientific developments, but we might have some ideas about how current trends will continue, beyond the general observation that important work will be driven by applications.

The safest bet is that there will be continuing progress on existing combinations of methods:  causal inference with rich models for potential outcomes, estimated using regularization; complex models for structured data such as networks evolving over time, robust inference for multilevel models; exploratory data analysis for overparameterized models (Mimno, Blei, and Engelhardt, 2015); subsetting and machine-learning meta-algorithms for different computational problems; and so forth.  In addition we expect progress on experimental design and sampling for structured data.

We can also be sure to see advances in computation.  From one direction, large and complex applied problems are being fit on faster computers, and we do not seem to have yet reached theoretical limits on efficiency of computational algorithms.  From the other direction, the availability of fast computation allows applied researchers to routinely do big computations, and this has direct impact on statistics research.  We have already seen this with hierarchical regressions, topic models, random forests, and deep nets, which have revolutionized many fields of application through their general availability.

Another general area that is ripe for development is model understanding, sometimes called interpretable machine learning (Murdoch et al., 2019, Molnar, 2020).  The paradox here is that the best way to understand a complicated model is often to approximate it with a simpler model, but then the question is, what is really being communicated here?  One potentially useful approach is to compute sensitivities of inferences to perturbations of data and model parameters (Giordano, Broderick, and Jordan, 2018), combining ideas of robustness and regularization with gradient-based computational methods that are used in many different statistical algorithms.

What are the biggest challenges and opportunities facing statisticians?  Three related trends in applications are big data, messy data, and complicated questions.  In some ways these trends go together:  when using data from more sources, it should be possible to make more more finely-grained inferences and decisions in problems ranging from personalized medicine to recommender systems to robot cars.

Does this mean that, as sample sizes get bigger and bigger, statistical inference will become less and less important than in the past, to the point that the machine-learning approach of purely predictive inference will replace the role of statistics except in some specialized ``small data'' applications?  We anticipate that no, there will always be a ``last mile problem'' by which researchers and decision makers will always be concerned with statistical issues of uncertainty and variation.  For example, machine learning methods can be used in drug discovery, and hierarchical differential equation models can be used in dosing models, but when estimating efficacy in the population, we think there is no way to avoid statistical issues of generalizing from sample to population, generalizing from treatment to control group, and generalizing from observed data to underlying constructs of interest.  This suggests to us that some of the most important statistical research of the next fifty years will lie at the interface of high-dimensional and nonparametric modeling and computation on one hand, and causal inference and decision making on the other.

A related question is what statistical ideas will be developed outside the area of statistics.  In the past twenty years, deep learning has had huge success, with traditional statistical theory often seeming to struggle to catch up.  Can we anticipate what new areas might arise, about which statisticians should become aware?  Much of the history of statistics can be viewed as the incorporation of ideas from outside.  Indeed, as a field we can count ourselves lucky that many of the new ideas of the past fifty years in topics as varied as causal inference, robustness, and exploratory data analysis were developed by researchers within statistics.  One strength of our field is its connection to applications, and to the extent that applied statistics or data science is now often done within applied fields of science and engineering, we can expect many of the new developments to come from there too, in the same way that earlier developments in statistics came from within applied fields such as psychology and genetics.  Statistics should continue to be open to ideas---general theoretical frameworks as well as specific models and methods---coming from other fields.


Finally, given that just about all new statistical and data science ideas are computationally expensive, we envision future research on validation of inferential methods, taking ideas such as unit testing from software engineering and applying them to problems of learning from noisy data. As our statistical methods become more advanced, there will a continuing need to understand the links between data, models, and substantive theory.

\section*{References}

\noindent

\bibitem Akaike, H. (1973).  Information theory and an extension of the maximum likelihood principle.  In {\em Proceedings of the Second International Symposium on Information Theory}, ed.\ B. N. Petrov and F. Csaki, 267--281.  Budapest:  Akademiai Kiado.  Reprinted in {\em Breakthroughs in Statistics}, ed.\ S. Kotz, 610--624.  New York: Springer (1992).

\bibitem Aldous, D. J. (1985). {\em Exchangeability and Related Topics}. Springer, Berlin.

\bibitem Amrhein, V., Greenland, S., and McShane, B. (2019).  Scientists rise up against statistical significance.  {\em Nature} {\bf 567}, 305--307.

\bibitem Anderlucci, L., Montanari, A., and Viroli, C. (2019). The importance of being clustered: Uncluttering the trends of statistics from 1970 to 2015. {\em Statistical Science} {\bf 34}, 280--300.

\bibitem Andrews, D. F., Bickel, P. J., Hampel, F. R., Huber, P. J., Rogers, W. H., and Tukey, J. W. (1972). {\em Robust Estimates of Location: Survey and Advances}.  Princeton University Press.

\bibitem Baron, R. M., and Kenny, D. A. (1986). The moderator–mediator variable distinction in social psychological research: Conceptual, strategic, and statistical considerations.  {\em Journal of Personality and Social Psychology} {\bf 51}, 1173--1182.

\bibitem Becker, R. A., Chambers, J. M., and Wilks, A. R. (1988).  {\em The New S Language:  A Programming Environment for Data Analysis and Graphics}.  Pacific Grove, Calif.:  Wads\-worth.

\bibitem Bengio, Y., LeCun, Y.,  and Hinton, G.  (2015). Deep learning. {\em Nature} {\bf 521}, 436--444.

\bibitem Benjamini, Y., and Hochberg, Y. (1995). Controlling the false discovery rate: A practical and powerful approach to multiple testing. {\em Journal of the Royal Statistical Society B} {\bf 57}, 289--300.

\bibitem Berger, J. O. (1985).  {\em Statistical Decision Theory and Bayesian Analysis}, second edition.  New York:  Springer.
  
\bibitem Bernardo, J. M., and Smith, A. F. M. (1994).  {\em Bayesian Theory}.  New York:  Wiley.

\bibitem Besag, J. (1974).  Spatial interaction and the statistical analysis of lattice systems (with discussion).  {\em Journal of the Royal Statistical Society B} {\bf 36}, 192--236.

\bibitem Besag, J. (1986).  On the statistical analysis of dirty pictures (with discussion).  {\em Journal of the Royal Statistical Society B} {\bf 48}, 259--302.

\bibitem Bland, J. M., and Altman, D. G. (1986). Statistical methods for assessing agreement between two methods of clinical measurement.  {\em Lancet} {\bf 327}, 307--310.
  
\bibitem Blei, D. M., Ng, A. Y., and Jordan, M. I. (2003).  Latent Dirichlet allocation.  {\em Journal of Machine Learning Research} {\bf 3}, 993--1022.

\bibitem Box, G. E. P. (1980).  Sampling and Bayes inference in scientific modelling and robustness.  {\em Journal of the Royal Statistical Society A} {\bf 143}, 383--430.


\bibitem Box, G. E. P., and Jenkins, G. M. (1976).  {\em Time Series Analysis:  Forecasting and Control}, second edition.  San Francisco: Holden-Day.



\bibitem Breiman, L. (2001).  Statistical modeling: The two cultures.  {\em Statistical Science} {\bf 16}, 199--231. 

\bibitem Breiman, L., Friedman, J. H., Olshen, R. A., and Stone, C. J. (1984). {\em Classification and Regression Trees}.  London: CRC Press.

\bibitem Brillinger, D. R. (1981).  {\em Time Series:  Data Analysis and Theory}, expanded edition.  San Francisco:  Holden-Day.

\bibitem Buntine, W. L., and Weigend, A. S. (1991).  Bayesian back-propagation.  {\em Complex Systems} {\bf 5}, 603--643.

\bibitem Candès, E. J., Romberg, J., and Tao, T. (2008). Robust uncertainty principles: Exact signal reconstruction from highly incomplete frequency information.  {\em IEEE Transactions on Information Theory} {\bf 52}, 489--509.

\bibitem Carvalho, C. M., Polson, N. G., and Scott, J. G. (2010). The horseshoe estimator for sparse signals.  {\em Biometrika} {\bf 97}, 465--480.

\bibitem Chambers, J. M. (1993). Greater or lesser statistics: A choice for future research. {\em Statistics and Computing} {\bf 3}, 182--184.

\bibitem Chambers, J. M., Cleveland, W. S., Kleiner, B., and Tukey, P. A. (1983).  {\em Graphical Methods for Data Analysis}.  Pacific Grove, Calif.:  Wadsworth.

\bibitem Chernozhukov, V., Chetverikov, D., Demirer, M., Duflo,  E., Hansen, C., Newey, W., and Robins, J. (2018). Double/debiased machine learning for treatment and structural parameters. {\em Econometrics Journal} {\bf 21}, C1--C68.

\bibitem Cleveland, W. S. (1985).  {\em The Elements of Graphing Data}. Monterey, Calif.:  Wads\-worth.

\bibitem Cortes, C., and Vapnik, V. (1995).  Support-vector networks.  {\em Machine Learning} {\bf 20}, 273--297.

\bibitem Cox, D. R. (1958).  {\em Planning of Experiments}.  New York:  Wiley.

\bibitem Cox, D. R. (1972). Regression models and life-tables.  {\em Journal of the Royal Statistical Society B} {\bf 34}, 187--220.


\bibitem Cronbach, L. J. (1975). Beyond the two disciplines of scientific psychology. {\em American Psychologist} {\bf 30}, 116--127.

\bibitem Del Moral, P. (1996). Nonlinear filtering: Interacting particle resolution. {\em Markov Processes and Related Fields} {\bf 2}, 555--580.

\bibitem Dempster, A. P., Laird, N. M., and Rubin, D. B. (1977).  Maximum likelihood from incomplete data via the EM algorithm (with discussion).  {\em Journal of the Royal Statistical Society B} {\bf 39}, 1--38.

\bibitem Dempster, A. P., Schatzoff, M., and Wermuth, N. (1977).  A simulation study of alternatives to ordinary least squares.  {\em Journal of the American Statistical Association} {\bf 72}, 77--91.

\bibitem Donoho, D. L. (1995).  De-noising by soft-thresholding.  {\em IEEE Transactions on Information Theory} {\bf 41}, 613--627.

\bibitem Donoho, D. L. (2006).  Compressed sensing.  {\em IEEE Transactions on Information Theory} {\bf 52},  1289--1306.

\bibitem Donoho, D. L. (2017).  50 years of data science.  {\em Journal of Computational and Graphical Statistics} {\bf 26}, 745--766.

\bibitem Donoho, D. L, and Johnstone, I. M. (1994).  Ideal spatial adaptation by wavelet shrinkage.  {\em Biometrika} {\bf 81}, 425--455.

\bibitem Duane, S., Kennedy, A. D., Pendleton, B. J., and Roweth, D. (1987).  Hybrid Monte Carlo.  {\em Physics Letters B} {\bf 195}, 216--222.

\bibitem Duncan, O. D. (1975).  {\em Introduction to Structural Equation Models}.  New York:  Academic Press.

\bibitem Efron, B. (1979).  Bootstrap methods:  Another look at the jackknife. {\em Annals of Statistics} {\bf 7}, 1--26.

\bibitem Efron, B. and Hastie, T. (2016).  {\em Computer Age Statistical Inference:  Algorithms, Evidence, and Data Science}.  Cambridge University Press.

\bibitem Efron, B., and Morris, C. (1971).  Limiting the risk of Bayes and empirical Bayes estimators---Part I:  The Bayes case.  {\em Journal of the American Statistical Association} {\bf 66}, 807--815.

\bibitem Efron, B., and Morris, C. (1972).  Limiting the risk of Bayes and empirical Bayes estimators---Part II:  The empirical Bayes case.  {\em Journal of the American Statistical Association} {\bf 67}, 130--139.

\bibitem Efron, B., and Tibshirani, R. J.  (1993).  {\em An Introduction to the Bootstrap}.  London:  Chapman and Hall.
  
\bibitem Fay, R. E., and Herriot, R. A. (1979). Estimates of income for small places:  An application of James-Stein
procedures to census data.  {\em Journal of the American Statistical Association} {\bf 74}, 269--277.

\bibitem Felsenstein, J. (1985). Confidence limits on phylogenies: An approach using the bootstrap.  {\em Evolution} {\bf 39}, 783--791.
  
\bibitem Ferguson, T. S. (1973). A Bayesian analysis of some nonparametric problems. {\em Annals of Statistics} {\bf 1}, 209--230.
  
\bibitem Freund, Y., and Schapire, R. E. (1997).  A decision-theoretic generalization of on-line learning and an application to boosting.  {\em Journal of Computer and System Sciences} {\bf 55}, 119--139.

\bibitem Friedman, J. H. (2001).  Greedy function approximation: A gradient boosting machine.  {\em Annals of Statistics} {\bf 29}, 1189--1232.

\bibitem Gabry, J., Simpson, D., Vehtari, A., Betancourt, M., and Gelman, A. (2019).  Visualization in Bayesian workflow (with discussion).  {\em Journal of the Royal Statistical Society A} {\bf 182}, 389--402.

\bibitem Geisser, S. (1975). The predictive sample reuse method with applications.  {\em Journal of the American Statistical Association} {\bf 70}, 320--328.

\bibitem Gelfand, A. E., and Smith, A. F. M. (1990).  Sampling-based approaches to calculating marginal densities.  {\em Journal of the
American Statistical Association} {\bf 85}, 398--409.

\bibitem Gelman, A. (2003).  A Bayesian formulation of exploratory data analysis and goodness-of-fit testing.  {\em International Statistical Review} {\bf 71}, 369--382.


\bibitem Gelman, A., Vehtari, A., Simpson, D., Margossian, C. C., Carpenter, B., Yao, Y., Bürkner, P. C., Kennedy, L., Gabry, J., and Modrák, M. (2020).  Bayesian workflow.\\ \url{www.stat.columbia.edu/~gelman/research/unpublished/Bayesian_Workflow_article.pdf}
  
\bibitem Geman, S., and Hwang, C. R. (1982).  Nonparametric maximum likelihood estimation by the method of sieves.  {\em Annals of Statistics} {\bf 10}, 401--414.

\bibitem Gigerenzer, G., and Todd, P. M. (1999).  {\em Simple Heuristics That Make Us Smart}.  Oxford University Press.

\bibitem Giordano, R., Broderick, T., and Jordan, M. I. (2018).  Covariances, robustness, and variational Bayes. {\em Journal of Machine Learning Research} {\bf 19}, 1--49.

\bibitem Good, I, J., and Gaskins, R. A. (1971).  Nonparametric roughness penalties for probability densities.
{\em Biometrika} {\bf 58}, 255--277.

\bibitem Goodfellow, I., Bengio, Y., and Courville, A. (2016). {\em Deep Learning}. Cambridge, Mass.:  MIT Press.

\bibitem Goodfellow, I., Pouget-Abadie, J., Mirza, M., Xu, B., Warde-Farley, D., Ozair, S., Courville, A., and Bengio, Y. (2014). Generative adversarial networks. Proceedings of the International Conference on Neural Information Processing Systems, 2672--2680.

\bibitem Gordon, N. J., Salmond, D. J., and Smith, A. F. M. (1993). Novel approach to nonlinear/non-Gaussian Bayesian state estimation. {\em IEE Proceedings F - Radar and Signal Processing} {\bf 140}, 107--113.

\bibitem Greenland, S. (2005).  Multiple-bias modelling for analysis of observational data.  {\em Journal of the Royal Statistical Society A} {\bf 168}, 267--306.

\bibitem Greenland, S., and Robins, J. M. (1986).  Identifiability, exchangeability, and epidemiological confounding.  {\em International Journal of Epidemiology} {\bf 15}, 413--419.

\bibitem Grenander, U. (1981).  {\em Abstract Inference}.  New York:  Wiley.

\bibitem Haavelmo, T. (1943).  The statistical implications of a system of simultaneous equations. {\em Econometrica} {\bf 11}, 1--12.

\bibitem Hastie, T., Tibshirani, R., and Wainwright, M. (2015). {\em Statistical Learning With Sparsity}.  London:  CRC Press.

\bibitem Heckerman, D., Geiger, D., and Chickering, D. M. (1995).  Learning Bayesian networks: The combination of knowledge and statistical data.  {\em Machine Learning} {\bf 20}, 197--243.

\bibitem Heckman, J. J., and Pinto, R. (2015).  Causal analysis after Haavelmo.  {\em Econometric Theory} {\bf 31}, 115--151.
  
\bibitem Henderson, C. R., Kempthorne, O., Searle, S. R., and von Krosigk, C. M. (1959).  The estimation of environmental and genetic trends from records subject to culling.  {\em Biometrics} {\bf 15}, 192--218.

\bibitem Heskes, T., Opper, M., Wiegerinck, W., Winther, O., and Zoeter, O. (2005). Approximate inference techniques with expectation constraints. {\em Journal of Statistical Mechanics: Theory and Experiment}, P11015.

\bibitem Hill, J. L. (2011).  Bayesian nonparametric modeling for causal inference. {\em Journal of Computational and Graphical Statistics} {\bf 20}, 217--240.

\bibitem Hinton, G. E., Osindero, S., and Teh, Y. W. (2006). A fast learning algorithm for deep belief nets.
{\em Neural Computation} {\bf 18}, 1527--1554.
  
\bibitem Hoeting, J., Madigan, D., Raftery, A. E., and Volinsky, C. (1999).  Bayesian model averaging (with discussion). {\em Statistical Science} {\bf 14}, 382--417.


\bibitem Huber, P. J. (1972).  Robust statistics:  A review.  {\em Annals of Mathematical Statistics} {\bf 43}, 1041--1067.

\bibitem Ihaka, R., and Gentleman, R. (1996).  R:  A language for data analysis and graphics.  {\em Journal of Computational and Graphical Statistics} {\bf 5}, 299--314.

\bibitem Imbens, G. W., and Angrist, J. D. (1994).  Identification and estimation of local average treatment effects.  {\em Econometrica} {\bf 62}, 467--475.

\bibitem Ioannidis, J. P. A. (2005).  Why most published research findings are false. {\em PLoS Medicine} {\bf 2} (8), e124.

\bibitem James, W., and Stein, C. (1960).  Estimation with quadratic loss.  In {\em Proceedings of the Fourth Berkeley Symposium} {\bf 1}, ed.\ J. Neyman, 361--380.  Berkeley:  University of California Press.

\bibitem Jordan, M., Ghahramani, Z., Jaakkola, T., and Saul, L. (1999). Introduction to variational methods for graphical models. {\em Machine Learning} {\bf 37}, 183--233.
  
\bibitem Kahneman, D., Slovic, P., and Tversky, A. (1982).  {\em Judgment
Under Uncertainty:  Heuristics and Biases}.
Cambridge University Press.

\bibitem Kimeldorf, G., and Wahba, G. (1971).  Some results on Tchebycheffian spline functions. {\em Journal of Mathematical Analysis and Applications} {\bf 33}, 82--95.
  
\bibitem Kitagawa, G. (1993). A Monte Carlo filtering and smoothing method for non-Gaussian nonlinear state space models. {\em Proceedings of the 2nd U.S.-Japan Joint Seminar on Statistical Time Series Analysis}, 110--131.

\bibitem Kolaczyk, E. D. (2009).  {\em Statistical Analysis of Network Data: Methods and Models}.  New York:  Springer.

\bibitem Kong, A., McCullagh, P., Meng, X. L., Nicolae, D., and Tan, Z. (2003).  A theory of statistical models for Monte Carlo integration (with discussion). {\em Journal of the Royal Statistical Society B} {\bf 65}, 585--618.

\bibitem Künsch, H. R. (1987).  Statistical aspects of self-similar processes.  {\em Proceedings of the First World Congress of the Bernoulli Society}, 67--74.

\bibitem Lavine, M. (1992). Some aspects of Polya tree distributions for statistical modelling. {\em Annals of Statistics} {\bf 20}, 1222--1235.

\bibitem Lax, J. R., and Phillips, J. H. (2012).  The democratic deficit in the states.  {\em American Journal of Political Science} {\bf 56}, 148--166.

\bibitem Lee, J. A. and Verleysen, M. (2007). {\em Nonlinear Dimensionality Reduction}. New York:  Springer.

\bibitem Liang, K. Y., and Zeger, S. L. (1986).  Longitudinal data analysis using generalized linear models.  {\em Biometrika} {\bf 73}, 13--22.

\bibitem Lindley, D. V., and Novick, M. R. (1981).  The role of exchangeability in inference.  {\em Annals of Statistics} {\bf 9}, 45--58.

\bibitem Lindley, D. V., and Smith, A. F. M. (1972).  Bayes estimates for the linear model.  {\em Journal of the Royal Statistical Society B} {\bf 34}, 1--41.

\bibitem Little, R. J. A. (1993).  Post-stratification:  A modeler's perspective.  {\em Journal of the American Statistical Association} {\bf 88}, 1001--1012.

\bibitem Little, R. J. A., and Rubin, D. B. (2002).  {\em Statistical Analysis with Missing Data}, second edition.  New York:  Wiley.

\bibitem Lunn, D., Spiegelhalter, D., Thomas, A., and Best, N. (2009). The BUGS project: Evolution, critique and future directions (with discussion). {\em Statistics in Medicine} {\bf 28}, 3049--3082.

\bibitem MacKay, D. J. C. (1992).  A practical Bayesian framework for backpropagation networks.  {\em Neural Computation} {\bf 4}, 448--472.

\bibitem Mallows, C. L. (1973).  Some comments on $C_p$.  {\em Technometrics} {\bf 15}, 661--675.

\bibitem Manski, C. F. (1990).  Nonparametric bounds on treatment effects.  {\em American Economic Review} {\bf 80}, 319--323.

\bibitem Marin, J. M., Pudlo, P., Robert, C. P., and Ryder, R. J. (2012). Approximate Bayesian computational methods. {\em Statistics and Computing} {\bf 22}, 1167--1180.

\bibitem Martin, G. M., Frazier, D. T., and Robert, C. P. (2020). Computing Bayes: Bayesian computation from 1763 to the 21st century. \url{arXiv:2004.06425}.

\bibitem Mauldin, R. D., Sudderth, W. D., and Williams, S. C. (1992). Polya trees and random distributions. {\em Annals of Statistics} {\bf 20}, 1203--1221.

\bibitem McCullagh, P., and Nelder, J. A. (1989).  {\em Generalized Linear Models}, second edition.  New York:  Chapman and Hall.

\bibitem Meehl, P. E.  (1978). Theoretical risks and tabular asterisks: Sir Karl, Sir Ronald, and the slow progress of soft psychology. {\em Journal of Consulting and Clinical Psychology} {\bf 46}, 806--834.

\bibitem Meng, X. L., and van Dyk, D. A.  (1997).  The EM algorithm---an
old folk-song sung to a fast new tune (with discussion).
{\em Journal of the Royal Statistical Society B} {\bf 59}, 511--567.

\bibitem Mimno, D., Blei, D. M., and Engelhardt, B. E. (2015).  Posterior predictive checks to quantify lack-of-fit in admixture models of latent population structure.  {\em Proceedings of the National Academy of Sciences} {\bf 112}, E3441--3450.

\bibitem Minka, T. (2001). Expectation propagation for approximate Bayesian inference. In {\em Proceedings of the Seventeenth Conference on Uncertainty in Artificial Intelligence}, ed.\ J. Breese and D. Koller, 362--369.

\bibitem Mockus, J. (1974).  The Bayes methods for seeking the extremal point.  {\em Kybernetes} {\bf 3}, 103--108.

\bibitem Mockus, J. (2012).  {\em Bayesian Approach to Global Optimization: Theory and Applications}.  Dordrecht:  Kluwer.
 
\bibitem Molnar, C. (2020). {\em Interpretable Machine Learning:  A Guide for Making Black Box Models Explainable}. \url{christophm.github.io/interpretable-ml-book}

\bibitem Morgan, S. L., and Winship, C. (2014). {\em Counterfactuals and Causal Inference: Methods and Principles for
Social Research}, second edition. Cambridge University Press.

\bibitem Müller, P., and Mitra, R. (2013). Bayesian nonparametric inference---why and how. {\em Bayesian Analysis}, {\bf 8}, 269--302.

\bibitem Murdoch, W. J., Singh, C., Kumbier, K., Abbasi-Asl, R., and Yu, B. (2019).  Definitions, methods, and applications in interpretable machine learning.  {\em Proceedings of the National Academy of Sciences} {\bf 116}, 22070--22080.

\bibitem Navarro, D. J. (2019).  Between the devil and the deep blue sea: Tensions between scientific judgement and statistical model selection. {\em Computational Brain and Behavior} {\bf 2}, 28--34.

\bibitem Neal, R. M. (1996).  {\em Bayesian Learning for Neural Networks}.  New York:  Springer.

\bibitem Nelder, J. A. (1977).  A reformulation of linear models (with discussion). {\em Journal of the Royal Statistical Society A} {\bf 140}, 48--76.

\bibitem Neyman, J. (1923). On the application of probability theory to agricultural experiments. Essay on principles. Section 9.  Translated and edited by D. M. Dabrowska and T. P. Speed. {\em Statistical Science} {\bf 5}, 463--480 (1990).

\bibitem Novick, M. R., Jackson, P. H., Thayer, D. T., and Cole, N. S. (1972).  Estimating multiple regressions in $m$ groups:  A cross validation study. {\em British Journal of Mathematical and Statistical Psychology} {\bf 25}, 33--50.

\bibitem O'Hagan, A. (1978). Curve fitting and optimal design for prediction (with discussion). {\em Journal of the Royal Statistical Society B} {\bf 40}, 1--42.

\bibitem Owen, A. B. (1988). Empirical likelihood ratio confidence intervals for a single functional.  {\em Biometrika} {\bf 75}, 237--249.

\bibitem Paatero, P. and Tapper, U. (1994). Positive matrix factorization: A non-negative factor model with optimal utilization of error estimates of data values. {\em Environmetrics} {\bf 5}, 111--126.
  
\bibitem Pearl, J. (2009).  {\em Causality}, second edition.  Cambridge University Press.

\bibitem Peters, J., Janzing, D., and Schölkopf, B. (2017).  {\em Elements of Causal Inference:  Foundations and Learning Algorithms}.  MIT Press.

\bibitem Pitman, J., and Yor, M. (1997). The two-parameter Poisson-Dirichlet distribution derived from a stable subordinator. {\em Annals of Probability} {\bf 25}, 855--900.

\bibitem Popper, K. R. (1957).  {\em The Poverty of Historicism}.  London:  Routledge and Kegan Paul.

\bibitem Pyro (2020).  Pyro:  Deep universal probabilistic programming.  \url{pyro.ai}

\bibitem Quenouille, M. H. (1949). Problems in plane sampling. {\em Annals of Mathematical Statistics} {\bf 20}, 355--375.

\bibitem Rasmussen, C. E., and Williams, C. K. I. (2006).  {\em Gaussian Processes for Machine Learning}.  Cambridge, Mass.:  MIT Press.

\bibitem Robbins, H. (1955).  An empirical Bayes approach to statistics.
In {\em Proceedings of the
Third Berkeley Symposium} {\bf 1}, ed.\ J. Neyman,
157--164.  Berkeley:  University
of California Press.

\bibitem Robbins, H. (1964).  The empirical Bayes approach to statistical
decision problems.  {\em Annals of Mathematical Statistics} {\bf 35}, 1--20.

\bibitem Rosenbaum, P. R., and Rubin, D. B. (1983). The central role of the propensity score in observational studies for causal effects.  {\em Biometrika} {\bf 70}, 41--55.

\bibitem Rubin, D. B. (1974).  Estimating causal effects of treatments in randomized and nonrandomized studies.  {\em Journal of Educational Psychology} {\bf 66}, 688--701.
  

\bibitem Rubin, D. B. (1984).  Bayesianly justifiable and relevant frequency calculations for the applied statistician.  {\em Annals of Statistics} {\bf 12}, 1151--1172.

\bibitem Rumelhart, D. E., Hinton, G. E., and Williams, R. J. (1987). Learning internal representations by error propagation. In {\em Parallel Distributed Processing:  Explorations in the Microstructure of Cognition: Foundations}, ed.\ Rumelhart, D. E. and McClelland, J. L., 318--362. Cambridge, Mass.: MIT Press.

\bibitem Savage, L. J. (1954).  {\em The Foundations of Statistics}.  New York:  Dover.

\bibitem Scheffé, H. (1959). {\em The Analysis of Variance}. New York: Wiley.

\bibitem Schmidhuber, J. (2015). Deep learning in neural networks: An overview. {\em Neural Networks} {\bf 61}, 85--117.

\bibitem Shahriari, B., Swersky, K., Wang, Z., Adams, R. P., and de Freitas, N. (2015). Taking the human out of the loop: A review of Bayesian optimization. {\em Proceedings of the IEEE} {\bf 104}, 148--175.

\bibitem Sheiner, L. B., Rosenberg, B., and Melmon, K. L. (1972).  Modelling of individual pharmacokinetics for computer-aided drug dosage.  {\em Computers and Biomedical Research} {\bf 5}, 441--459.

\bibitem Shen, X., and Wong, W. H. (1994).  Convergence rate of sieve estimates.  {\em Annals of Statistics} {\bf 22}, 580--615.

\bibitem Silver, D., Schrittwieser, J.,  Simonyan, K., Antonoglou, I., Huang, A., Guez, A., Hubert, T., Baker, L., Lai, M., Bolton, A., Chen, Y., Lillicrap, T., Hui, F., Sifre, L., van den Driessche, G., Graepel, T., and Hassabis, D. (2017).  Mastering the game of Go without human knowledge.  {\em Nature} {\bf 550}, 354--359. 

\bibitem Simmons, J., Nelson, L., and Simonsohn, U. (2011).  False-positive psychology:  Undisclosed flexibility in data collection and analysis allow presenting anything as significant.  {\em Psychological Science} {\bf 22}, 1359--1366.


\bibitem Spiegelhalter, D., Thomas, A., Best, N., Gilks, W., and Lunn, D. (1994). BUGS:  Bayesian inference using Gibbs sampling.  MRC Biostatistics Unit, Cambridge, England.\\
\url{www.mrc-bsu.cam.ac.uk/bugs}

\bibitem Spirtes, P., Glymour C., and Scheines, R. (1993).  {\em Causation, Prediction, and Search}. New York:  Springer

\bibitem Stan Development Team (2020). Stan modeling language users guide and reference manual, version 2.25.  \url{mc-stan.org}

\bibitem Stein, C. (1955).  Inadmissibility of the usual estimator for the mean of a multivariate normal distribution.  In {\em Proceedings of the Third Berkeley Symposium} {\bf 1}, ed.\ J. Neyman, 197--206.  Berkeley:  University of California Press.


\bibitem Stigler, S. M. (1986).  {\em The History of Statistics}.  Cambridge,
Mass.:  Harvard University Press.
  
\bibitem Stigler, S. M. (2010).  The changing history of robustness.  {\em American Statistician} {\bf 64}, 277--281.

\bibitem Stigler, S. M. (2016). {\em The Seven Pillars of Statistical Wisdom}. Cambridge, Mass.: Harvard University Press.

\bibitem Stone, M. (1974).  Cross-validatory choice and assessment of statistical predictions (with discussion).
  {\em Journal of the Royal Statistical Society B} {\bf 36}, 111--147.

\bibitem Sutton, R. S., and Barto, A. G. (2018).  {\em Reinforcement Learning:  An Introduction}, second edition.  Cambridge, Mass.: MIT Press.

\bibitem Tavaré, S, Balding, D. J., Griffiths, R. C., and Donnelly, P. (1997). Inferring coalescence times from DNA sequence data. {\em Genetics} {\bf 145}, 505--518.

\bibitem Tensorflow (2000).  Tensorflow:  An end-to-end open source machine learning platform.\\ \url{www.tensorflow.org}

\bibitem Tibshirani, R. (1996). Regression shrinkage and selection via the lasso. {\em Journal of the Royal Statistical Society B} {\bf 58}, 267--288.

\bibitem Tufte, E. R. (1983).  {\em The Visual Display of Quantitative Information}.  Cheshire, Conn.: Graphics Press.

\bibitem Tukey, J. W. (1953).  {\em The Problem of Multiple Comparisons}.  Unpublished manuscript.

\bibitem Tukey, J. W. (1958). Bias and confidence in not quite large samples (abstract). {\em Annals of Mathematical Statistics} {\bf 29}, 614.

\bibitem Tukey, J. W. (1960).  A survey of sampling from contaminated distributions.  In {\em Contributions to Probability and Statistics:  Essays in Honor of Harold Hotelling}, ed.\ I. Olkin, S. G. Ghurye, W. Hoeffding, W. G. Madow, and H. B. Mann, 448--485.  Stanford University Press.

\bibitem Tukey, J. W. (1962).  The future of data analysis.   {\em Annals of Mathematical Statistics} {\bf 33}, 1--67.

\bibitem Tukey, J. W. (1977).  {\em Exploratory Data Analysis}.  Reading, Mass.: Addison-Wesley.

\bibitem Unwin, A., Volinsky, C., and Winkler, S. (2003). Parallel coordinates for exploratory modelling analysis. {\em Computational Statistics and Data Analysis} {\bf 43}, 553--564.

\bibitem VanderWeele, T. J. (2015).  {\em Explanation in Causal Inference: Methods for Mediation and Interaction}.  Cambridge University Press.

\bibitem van Zwet, E., Schwab, S., and Senn, S. (2020).  The statistical properties of RCTs and a proposal for shrinkage.  \url{arxiv.org/abs/2011.15004}
  
\bibitem Vapnik, V. N. (1998).  {\em Statistical Learning Theory}.  New York:  Wiley.

\bibitem Wager, S., and Athey, S. (2018).  Estimation and inference of heterogeneous treatment effects using random forests. {\em Journal of the American Statistical Association} {\bf 113}, 1228--1242.

\bibitem Wahba, G. (1978).  Improper priors, spline smoothing and the problem of guarding against model errors in regression. {\em Journal of the Royal Statistical Society B} {\bf 40}, 364--372.

\bibitem Wahba, G. (2002).  Soft and hard classification by reproducing kernel Hilbert space methods.  {\em Proceedings of the National Academy of Sciences} {\bf 99}, 16524--16530.

\bibitem Wahba, G., and Wold, S. (1975). A completely automatic French curve: Fitting spline functions by
cross-validation. {\em Communications in Statistics} {\bf 4}, 1--17.

\bibitem Wald, A. (1949).  Statistical decision functions.  {\em Annals of Mathematical Statistics} {\bf 20}, 165--205.

\bibitem Watanabe, S. (2010).  Asymptotic equivalence of Bayes cross validation and widely applicable information criterion in singular learning theory. {\em Journal of Machine Learning Research} {\bf 11}, 3571--3594.

\bibitem Welch, B. L. (1937).  On the z-test in randomized blocks and latin squares.  {\em Biometrika} {\bf 29}, 21--52.

\bibitem Werbos, P. J. (1981). Applications of advances in nonlinear sensitivity analysis. {\em Proceedings of the 10th IFIP Conference}, 762--770.

\bibitem Wermouth, N. (1980). Linear recursive equations, covariance selection, and path analysis.  {\em Journal of the American Statistical Association} {\bf 75}, 963--972.

\bibitem White, H. (1980). A heteroskedasticity-consistent covariance matrix estimator and a direct test for heteroskedasticity.  {\em Econometrica} {\bf 48}, 817--838.

\bibitem Wickham, H. (2006). Exploratory model analysis with R and GGobi.\\ \url{had.co.nz/model-vis/2007-jsm.pdf}

\bibitem Wickham, H. (2016). {\em ggplot2: Elegant Graphics for Data Analysis}. New York:  Springer. 
  
\bibitem Wilkinson, L. (2005).  {\em The Grammar of Graphics}, second edition.  New York:  Springer.

\bibitem Wold, H. O. A. (1954).  Causality and econometrics.  {\em Econometrica} {\bf 22}, 162--177.

\bibitem Wolpert, D. H. (1992).  Stacked generalization. {\em Neural Networks} {\bf 5}, 241--259.

\bibitem Wright, S. (1923).  The theory of path coefficients:  A reply to Niles' criticism.  {\em Genetics} {\bf 8}, 239--255.

\bibitem Wu, Y. N., Guo, C. E., and Zhu, S. C. (2004). Perceptual scaling. In {\em Applied Bayesian Modeling and Causal Inference from an Incomplete Data Perspective}, ed.\ A. Gelman and X. L. Meng.  New York:  Wiley.

\bibitem Yu, B. (2013).  Stability.  {\em Bernoulli} {\bf 19}, 1484--1500.


\end{document}